\documentclass[aip,reprint,twocolumn]{revtex4-2}

\usepackage{graphicx}% Include figure files
\usepackage{dcolumn}% Align table columns on decimal point
\usepackage{bm}% bold math
\usepackage{color}
\usepackage{amsmath}
\usepackage{amssymb}
\usepackage{graphicx}
\usepackage[unicode=true,pdfusetitle,
 bookmarks=true,bookmarksnumbered=false,bookmarksopen=false,
 breaklinks=false,pdfborder={0 0 0},pdfborderstyle={},backref=false,colorlinks=true]
 {hyperref}
\hypersetup{
 pdfborderstyle={},citecolor=blue}

\usepackage[utf8]{inputenc}
\usepackage[T1]{fontenc}
\usepackage{mathptmx}

\begin{document}

 \title{Influence of time-delayed feedback on the dynamics of temporal localized structures in passively mode-locked semiconductor lasers}
 \author{Thomas G. Seidel}
 \affiliation{Institute for Theoretical Physics, University of M\"unster, Wilhelm-Klemm-Str. 9, 48149 M\"unster, Germany}
\author{Julien Javaloyes}
\affiliation{Departament de Física \& IAC-3, Universitat de les Illes Balears, C/ Valldemossa km 7.5, 07122 Mallorca, Spain}
\author{Svetlana V. Gurevich}
\email{gurevics@uni-muenster.de}
\affiliation{Institute for Theoretical Physics, University of M\"unster, Wilhelm-Klemm-Str. 9, 48149 M\"unster, Germany}
 \affiliation{Center for Nonlinear Science (CeNoS), University of M\"unster, Corrensstrasse 2, 48149 M\"unster, Germany}
 \affiliation{Departament de Física \& IAC-3, Universitat de les Illes Balears,
 C/ Valldemossa km 7.5, 07122 Mallorca, Spain}
\begin{abstract}
In this paper, we analyze the effect of optical feedback on the dynamics of
a passively mode-locked ring laser operating in the regime of temporal localized structures. This laser system is modeled by a system of delay differential equations, which include delay terms associated with the laser cavity and the feedback loop.
Using a combination of direct numerical simulations and path-continuation techniques, we show that the feedback loop creates echos of the main pulse whose position and size strongly depend on the feedback parameters. We demonstrate that in the long-cavity regime, these echos can successively replace the main pulses, which defines their lifetime. This pulse instability mechanism originates from a global bifurcation of the saddle-node infinite-period type. In addition, we show that, under the influence of noise, the stable pulses exhibit forms of behavior characteristic of excitable systems. Furthermore, for the  harmonic solutions consisting of multiple equispaced pulses per round-trip we show that if the location of the pulses coincide with the echo of another, the range of stability of these solutions is increased. Finally, it is shown that around these resonances, branches of different solutions are connected by period doubling bifurcations.
\end{abstract}

\maketitle
\textbf{We theoretically analyze the effect of time-delayed feedback on the dynamics of a passively mode-locked ring laser model consisting of a set of coupled delay differential equations in the long cavity regime. By conducting extended direct numerical simulations and utilizing path-continuation techniques we find that the presence of a time-delayed feedback loop yields a wide variety of complex dynamics in the model including a global instability, excitability and period doubling bifurcations. In addition, we show that the stability and dynamics of harmonic solutions, consisting of different numbers of pulses per round-trip is critically influenced by the feedback parameters, namely the feedback delay time and the feedback rate.}

\section{Introduction}

Passive mode locking (PML) is a well-known method for achieving ultra short optical pulses with high repetition rates~\cite{haus00rev}. PML is a subject of intense research due to a wide variety of applications including optical data communication, optical clocking, metrology, medical imaging and surgery~\cite{keller96,lorenser04,UHH-NAT-02,Keller2003,AJ-BOOK-17}. Semiconductor mode-locked lasers are of particular interest, as they are relatively easy to manufacture and have a small footprint. Among the latter, vertical cavity surface-emitting semiconductor lasers (VCSELs) are prominent laser sources allowing to obtain lasing with high output power and excellent beam quality~\cite{WMD-OL-08,Innoptics,DMB-JSTQE-13}. Here, the PML  can be archived  by closing the external cavity with a semiconductor saturable absorber (SA) mirror~\cite{haus00rev,AJ-BOOK-17}. Recently, a regime of temporal localization allowing arbitrary low repetition rates was demonstrated in such systems \cite{MJB-PRL-14,CJM-PRA-16,JCM-PRL-16,CSV-OL-18, SJG-PRA-18,SCM-PRL-19,SHJ-PRAp-20}. In this regime in which the cavity round trip is much larger than the gain recovery time, the PML pulses become individually addressable \emph{temporal localized states} (TLSs) coexisting with the off solution~\cite{LCK-NAP-10,HBJ_NP_13,MJB-PRL-14}.  In this long cavity regime, although the PML pulses have a duration $\sim$1~ps, they leave a material trail in the gain medium that follows their emission. As the gain recovery $\sim$1~ns is the slowest variable, it defines the effective duration of the TLS, making the latter a stiff multiple timescale object and their theoretical and numerical analysis tedious~\cite{SJG-PRA-18, SJG-OL-18}.

The rich nonlinear PML dynamics in the short cavity limit  can be controlled with, e.g., time-delayed feedback (TDF)~\cite{AAV-APB-15, JNS-PRE-16, Jaurigue2017,Bartolo2021} or coherent optical injection~\cite{AHP-JOSAB-16}. In particular, TDF in mode-locked laser can either occur from parasitic reflections from intra-cavity lenses but it can also be applied on purpose to control the output of the laser, e.g., improving the time jitter in high repetition rate mode-locked lasers or for controlling the pulse train repetition rate~\cite{Otto2012,Jaurigue2017,AAV-APB-15}. However, the impact of TDF in PML semiconductor lasers operated in the long cavity limit still remains poorly understood. Recently, the impact of TDF on the dynamics of a VCSEL operating in the TLS regime was studied both experimentally and theoretically in Ref.~\cite{Bartolo2021}. In particular, it was shown that TDF can be used as a solution selector that either reinforces or hinders the appearance of one of the multistable harmonic arrangements of equispaced pulses. These experimental observations were confirmed theoretically using a generic delay differential equation (DDE) model for a unidirectionally emitting ring PML laser. This allowed to evidence asymmetrical resonance tongues occurring due to the parity symmetry-breaking induced by the gain depletion~\cite{JCM-PRL-16}.

In this paper we build upon this work~\cite{Bartolo2021} by presenting an
in-depth bifurcation analysis of the PML dynamics in the TLS regime under the influence of time-delayed optical feedback. Using a combination of direct numerical simulations and path-continuation techniques we shall show that the feedback loop creates echos of the main pulse which position and size strongly depend on the TDF parameters. When the echo is placed sufficiently close to the left side of the main pulse, TDF
can lead to an instability while increasing the gain. We demonstrate that in the long-cavity regime, this instability originates from a global bifurcation of the saddle-node infinite-period type, and we explain why the amplitude exhibit behaviors characteristic of excitable systems. In addition,  for the harmonic solutions consisting of multiple pulses per round-trip we show that applying resonant feedback, i.e., if  the location of a pulse coincide with an echo, the range of stability of this solution is increased. Finally, around the resonance, branches of different solutions are connected by period doubling bifurcations.

\section{Model system}

\begin{figure}
	\centering
	\includegraphics[width=1\columnwidth]{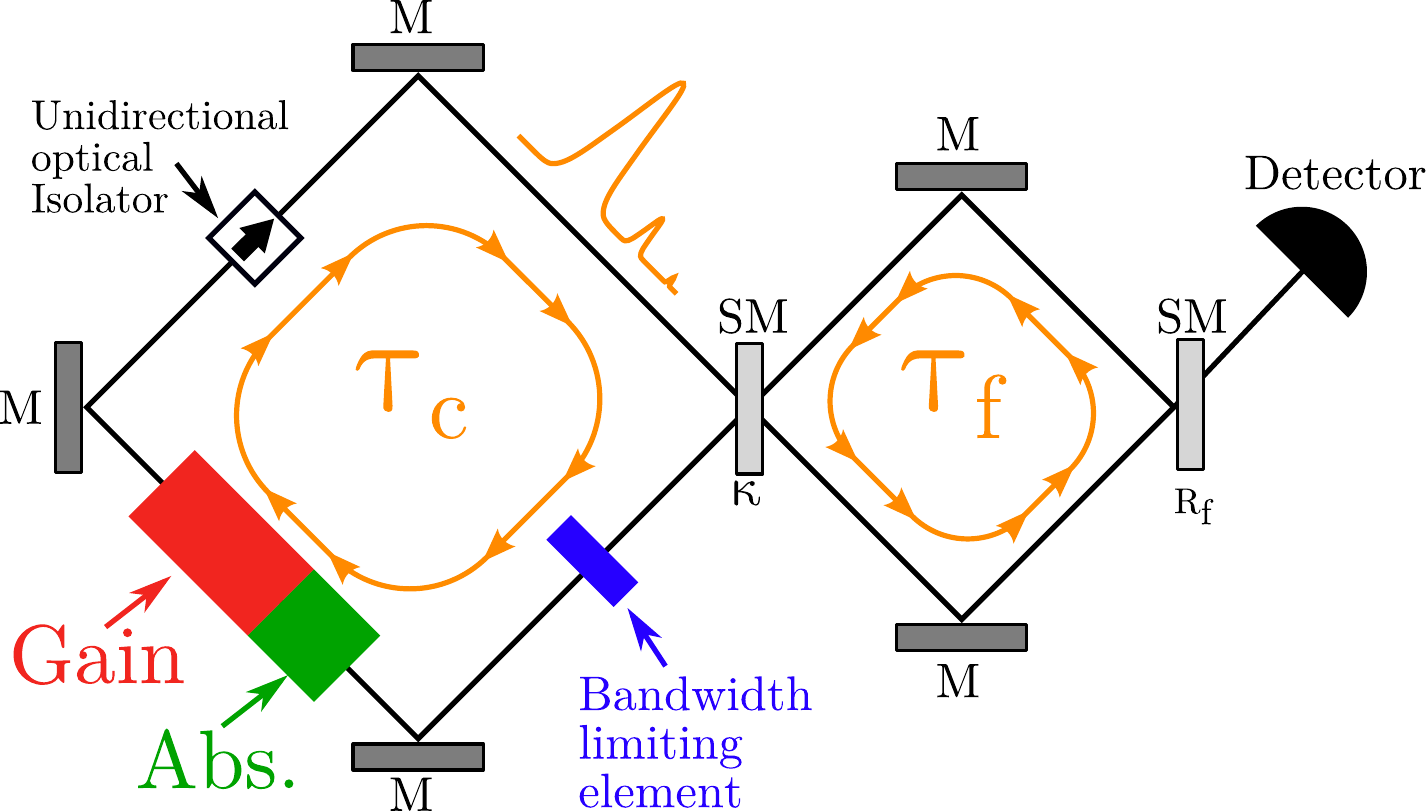}
	\caption{A schematic of the experimental setup with two ring cavities of length $\tau_c$ and $\tau_f$. M stands for mirror while SM are semitransparent mirrors with the respective reflectivities $\kappa$ and $R_f$, respectively.}
	\label{fig:setup}
\end{figure}
A sketch of a PML laser with an external feedback loop is shown in Fig.~\ref{fig:setup}. It consists of two ring cavities with round-trip times $\tau_c$ and $\tau_f$, respectively,  confined by mirrors (M) with 100$\,\%$ reflectivity. In the main cavity of the round-trip $\tau_c$, a gain and a saturable absorber sections are placed. The unidirectional optical isolator ensures that the pulses only propagate in one direction. The bandwidth limiting element ensures that only certain laser modes are amplified while others are suppressed. The two cavities are connected by a semitransparent mirror (SM) which reflects a fraction of $\kappa$ of the electric field. In the second cavity $\tau_f$ a partially reflective mirror with reflectivity $R_f$ is placed, i.e., a fraction of $\eta=(1-\kappa)\sqrt{R_f}$ of the electric field is fed back into the main cavity after a time delay $\tau_f$. The rest of the optical power leaves the coupled cavity setup towards, e.g., a photodetector.

The experimental work in Ref.~\cite{Bartolo2021} consisted in two micro-cavities coupled face-to-face. The latter can be modelled using the first principle approach of Ref.~\cite{MB-JQE-05,SCM-PRL-19} which naturally incorporate the dispersive effects of the micro-cavities. However, in order to better concentrate on the most generic instability mechanisms and to complement the theoretical analysis in Ref.~\cite{Bartolo2021}, we employ the model developed in Ref.~\cite{VT-PRA-05} that corresponds to a PML laser in a unidirectional ring geometry. The latter is extended by an additional optical feedback term~\cite{Otto2012,AAV-APB-15,JPR-PRA-15,Seidel2020}. The equations for the field amplitude $A$, the gain $G$, and the absorption $Q$ read	
\begin{align}
\frac{\dot{A}}{\gamma} = & \sqrt{\kappa} \exp\left[\frac{1-i\alpha_g}{2}G\left(t-\tau_{c}\right)-\frac{1-i\alpha_a}{2}Q\left(t-\tau_{c}\right)\right] \nonumber \\
& \times A\left(t-\tau_{c}\right)-A\left(t\right)+\eta e^{i\Omega}A\left(t-\tau_{f}\right),\label{eq:sys_eq_A}\\
\dot{G} = & g_{0}-\Gamma \cdot G-e^{-Q}\left(e^{G}-1\right)\left|A\right|^{2},\label{eq:sys_eq_G}\\
\dot{Q} = & q_{0}-Q-s\left(1-e^{-Q}\right)\left|A\right|^{2}\,.\label{eq:sys_eq_Q}
\end{align}
Here, time is normalized to the SA recovery time, $g_0$ is the  pumping  rate, $\Gamma$ is the gain recovery rate and $q_0$ is the  value of the unsaturated losses that determines the modulation depth of the SA. Further, $\alpha_g$ and $\alpha_a$ represent the linewidth enhancement factors of the gain and absorber sections, respectively, $s$ is the ratio of the saturation energy of the gain and of the SA sections, $\gamma$ is the bandwidth of the spectral filter and $\Omega$ is a feedback phase. The spatial boundary condition due to the closing of a cavity onto itself after a propagation length appears as a time delay $\tau_c$. The latter governs the fundamental repetition rate of the PML laser. In the long cavity limit, the value of $\Omega$ is irrelevant since a infinitesimal frequency shift allows canceling a nonzero value of $\Omega$. 
In this situation, the lasing threshold, determined by the value where the off solution $(A,G,Q)=(0,g_0,q_0)$ becomes linearly unstable, reads $g_{0,\text{th}}=~\Gamma \left[q_0-\ln \kappa + 2 \ln(1-\eta)\right].$
We define a normalized gain as $g=g_0/g_{0,\text{th}}$ and, if not stated differently, we fix the following parameters: $(\gamma,\kappa,\Gamma,q_0,s,\Omega)=(10,0.8,0.04,0.3,10,0)$ while the cavity round-trip is $\tau_c=100$.
\begin{figure}
	\centering
	\includegraphics[width=1\columnwidth]{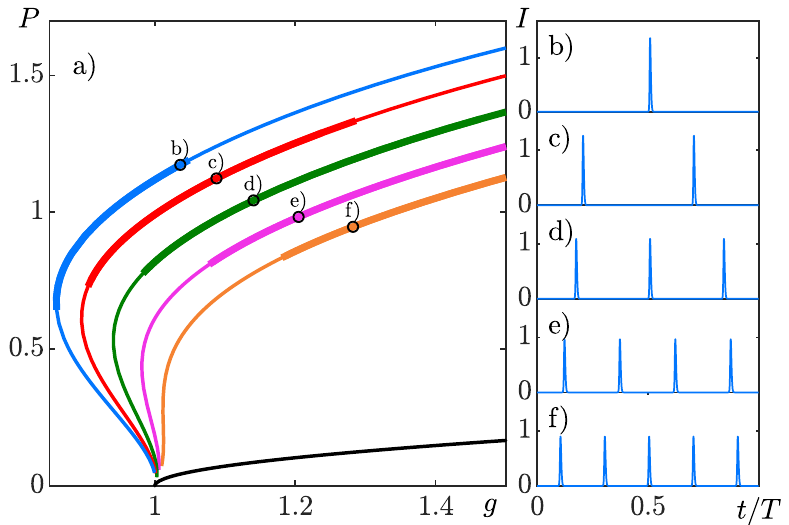}
	\caption{a) Path-continuation of the FML (blue) and HML$_{2-5}$ (red, green, pink, yellow) solutions of Eqs.~\eqref{eq:sys_eq_A}-\eqref{eq:sys_eq_Q} as a function of the normalized gain $g$ in the long delay limit for $\eta=0$. The value of $P=\max\left(|A|\right)$ is shown. Thick and thin lines denote stable and unstable solutions, respectively. The black line at the bottom corresponds to the CW branch. Its stability is omitted in the plot. The parameters are $(\alpha_g,\,\alpha_a)=(0.8,\,0.5)$. Panels b)-f) show the solution intensity profiles $I=|A|^2$ over a period $T$ at the position marked by the black circles in a).}
	\label{fig:branch_harmonic}
\end{figure}
Temporal localized structures appear in the time-delayed system~\eqref{eq:sys_eq_A}-\eqref{eq:sys_eq_Q} in the long-delay limit as periodic orbits whose period $T$ is always slightly larger than the time delay~\cite{MJB-PRL-14,MJC-JSTQE-15,SJG-PRA-18}. In particular, in the absence of the additional feedback loop, i.e., $\eta=0$, at $g=1$ a branch of continuous wave (CW) solutions emerges, see the black line in Fig.~\ref{fig:branch_harmonic}~(a). For the short delays, this branch undergoes a supercritical Andronov-Hopf (AH) bifurcation and a branch of a fundamental PML solution (FML) appears exhibiting a single pulse per round-trip~\cite{VTK-OL-04} as seen in Fig.~\ref{fig:branch_harmonic}~(b). As the pump $g_0$ or the cavity length $\tau_c$ is increased, the so-called regimes of harmonic mode-locking (HML$_n$), that consists in $n$ equispaced pulses per round-trip, develops at the expense of the fundamental PML solution, cf. Fig.~\ref{fig:branch_harmonic}~(b-f). Note that the intensity of the HML$_n$ solution decreases with increasing $n$ because the available population inversion must be distributed between a larger number of pulses. However, by increasing $\tau_c$, the AH bifurcations become subcritical, and the emanating periodic branches can extend below the lasing threshold $g_{0,\text{th}}$ and coexist with the off state, which defines the TLS regime. There, the TLSs gain stability in a saddle-node bifurcation of limit cycles (SN) for the FML solution or a torus bifurcation for HML$_n$ solutions, see Fig.~\ref{fig:branch_harmonic}.
% There, the pulsed solutions appear trough saddle-node bifurcations of limit cycles (SN), see Fig.~\ref{fig:branch_harmonic}.
% for the FML solution or a torus bifurcation for HML$_n$ solutions
% JJ: The HML are stable at the fold no ? At least they look that way.

\section{Influence of the time-delayed feedback}
\begin{figure}
	\centering
	\includegraphics[width=1\columnwidth]{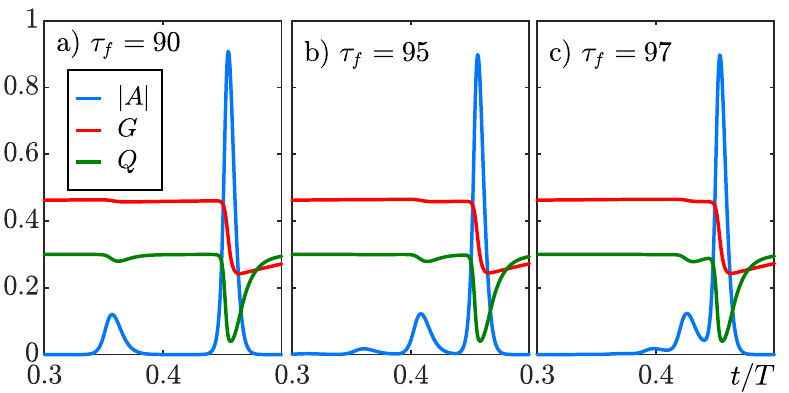}
	\caption{Direct numerical simulations of the single TLS of Eqs.~\eqref{eq:sys_eq_A}-\eqref{eq:sys_eq_Q} for different values the feedback delay: a) $\tau_f=90$, b) $\tau_f=95$, c) $\tau_f=97$ while the round-trip in the main cavity is $\tau_c=100$. Other parameters are $(\alpha_g,\alpha_a,\eta,g)=(0.6,0.5,0.006,0.921)$. The temporal separation between the echo and the main pulse is $\Delta \tau = \tau_c-\tau_f$.}
	\label{fig:DNS}
\end{figure}

\begin{figure}
	\centering
	\includegraphics[width=1\columnwidth]{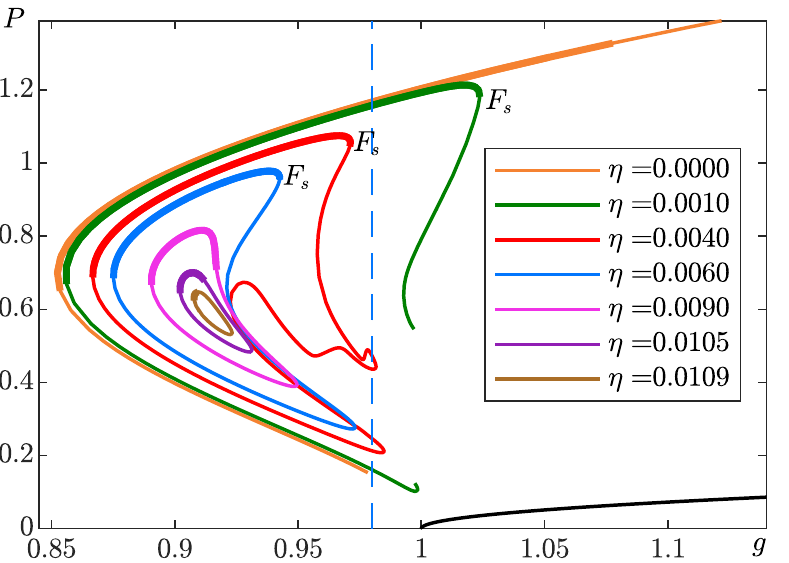}
	\caption{Bifurcation diagrams of Eqs. (\ref{eq:sys_eq_A})-(\ref{eq:sys_eq_Q}) in $(g,\,P)$ plane of a single TLS solution for different values of $\eta$. Thick and thin lines denote stable and unstable solutions, respectively. The thin black line in the bottom right is the CW branch for $\eta=0.001$. Its stability is omitted in the plot. The solution profile along each branch can be found as an animation in the supplementary material. Larger $\eta$ decrease the range of stability of the TLSs. Dashed vertical line indicates the $g$ value where numerical simulations in Fig.~\ref{fig:DNS SNIPER} were carried out. Parameters are  $(\alpha_g,\alpha_a,\tau_c,\tau_f)=(0,0,100,80)$.}
	\label{fig:branches_different_eta}
\end{figure}

%We define $P=\max\left(|A|\right)$ and $I = \max\left(\left|A\right|^2\right)$ as the norms used for plotting.
% 
% 
For $\eta\neq0$, the feedback loop creates echos of the main pulse~\cite{Otto2012, Bartolo2021} whose timing can directly be controlled via $\Delta \tau = \tau_f-\tau_c$, whereas their amplitude is determined by the feedback rate $\eta$. This is visualized in Fig.~\ref{fig:DNS} where direct numerical simulations of the FML solution of Eqs.~\eqref{eq:sys_eq_A}-\eqref{eq:sys_eq_Q} for fixed $\eta$ and different values of $\tau_f$ are performed. The echo pulses move closer and closer towards the main pulse as $\Delta \tau$ approaches zero. In Fig.~\ref{fig:DNS}~(b),(c) one can also see the effect of second order feedback, i.e., the echo's echo.
Figure~\ref{fig:branches_different_eta} shows the branches of the single TLS in the $(g, P)$ plane obtained within DDE-BIFTOOL~\cite{DDEBT} for the increasing values of the feedback rate $\eta$. One can see that for $\eta\neq0$ the branches experience an additional fold $F_s$ on the hight power part of the branch and the branches form closed loops. The loops become smaller with increasing $\eta$ and the stability terminates at $F_s$ for the weak feedback (cf. $\eta=0.006,0.004,0.001$ in Fig.~\ref{fig:branches_different_eta}) while on the other branches a AH bifurcation occurs. Note that the branches for $\eta=0$ and $\eta=0.001$ end abruptly due to numerical problems during their continuation.
As mentioned above, the amplitude of the echos induced by the feedback is determined by the feedback rate $\eta$. 
An echo located on the left of the main pulse depletes the gain at the expense of the latter.
Consequently, if an echo is located too close to the main pulse, the gain cannot recover fast enough before the main pulse arrives which may lead to an instability for the main pulse. The depletion of the gain depends on the amplitude of the echo which ultimately is proportional to $\eta$. It explains the results of Fig.~\ref{fig:branches_different_eta} in which the larger $\eta$, the smaller the stability range.

\section{Global bifurcation}

\begin{figure}
	\centering
	\includegraphics[width=1\columnwidth]{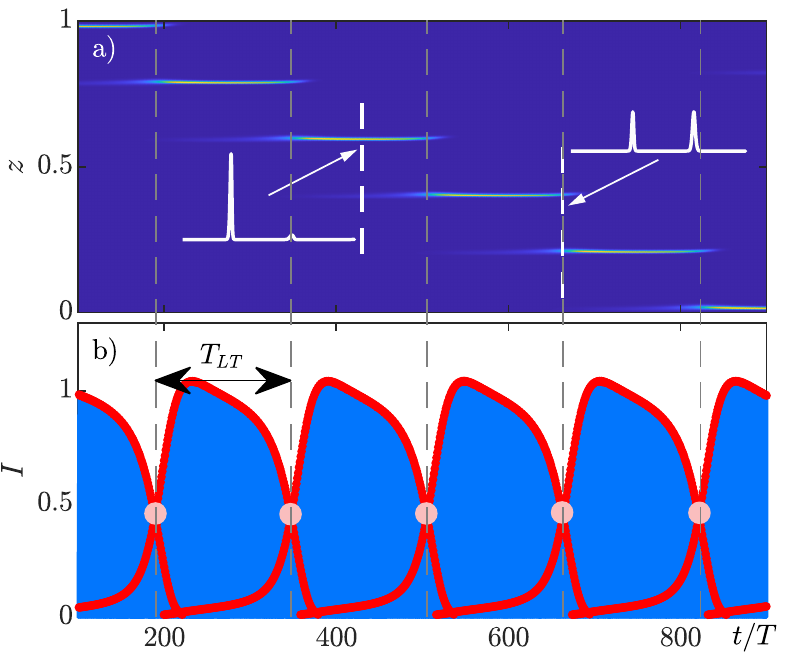}
	\caption{Direct numerical simulations of a single TLS of Eqs.~(\ref{eq:sys_eq_A})-(\ref{eq:sys_eq_Q}) for $\eta=0.006$ and $g=0.98$. In a) the two-time representation and b) time trace of the pulse intensity $I=|A|^2$ are presented, respectively. The simulation is carried out above the bifurcation point $F_s$ as indicated by the blue dashed line in Fig.~\ref{fig:branches_different_eta}. The insets in a) show the solution profiles along the white dashed lines. Red dots in b) indicate the maximal intensity in each round-trip and the pink dots as well as gray dashed lines indicate the time intervals defining pulse lifetime $T_{LT}$.}
	\label{fig:DNS SNIPER}
\end{figure}

\begin{figure}
	\centering
	\includegraphics[width=1\columnwidth]{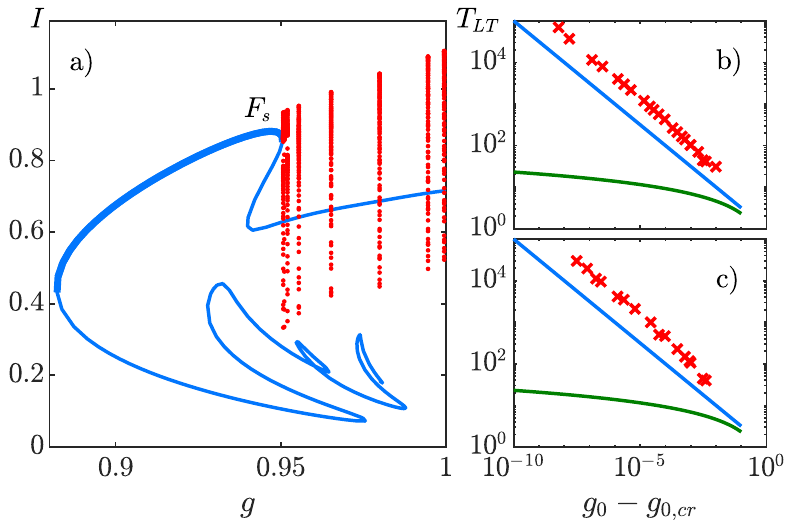}
	\caption{a) Bifurcation diagram of the single TLS solution of Eqs.~(\ref{eq:sys_eq_A})-(\ref{eq:sys_eq_Q}) for $(\alpha_g,\alpha_a,\eta,\tau_c,\tau_f)=(0.8,0.5,0.006,100,80)$. The blue branch is obtained from the path-continuation, while the red dots are results from direct numerical simulations and stand for the maximum intensity in each round-trip, cf. red dots in Fig.~\ref{fig:DNS SNIPER} (b). b),c) Lifetime $T_{LT}$ as a function of the distance to the bifurcation point $g_{0,cr}$ where the global bifurcation ($F_s$ in (a)) and Fig.~(\ref{fig:branches_different_eta})) occurs for b) $(\alpha_g,\alpha_a)=(0.01,0.0)$ and c) $(\alpha_g,\alpha_a)=(0.8,0.5)$. The lines correspond to the theoretically predicted scaling behavior for a homoclinic (green) as well as for a SNIPER (blue) bifurcations.}
	\label{fig:branch_sniper}
\end{figure}

The natural follow-up question is to wonder what happens to the TLSs above the bifurcation point $F_s$. To clarify this issue, direct numerical simulations for fixed $\eta$ at the $g$ value indicated by the blue dashed line in Fig.~\ref{fig:branches_different_eta} were conducted. They indicate that in this regime the main pulse  performs "jumps" within the cavity towards the position where its echo was previously located. This replacement process is visualized in Fig.~\ref{fig:DNS SNIPER}, where a two-time representation~\cite{GP-PRL-96,FG-PRE-20} and the corresponding time trace are shown in Fig.~\ref{fig:DNS SNIPER}~(a),(b), respectively. For the two-time representation, we use that the dynamics of the TLSs can be separated into two time scales. The fast time scale $z$ governs the dynamics within one period, and the slow scale describes the dynamics from one round-trip to the next one. Looking at Fig. \ref{fig:DNS SNIPER}, one can see that above $F_s$, the echo depletes the gain which cannot recover by the time the main pulse arrives. Since there is not enough amplification left, the main pulse slowly shrinks while the echo grows. At a certain point the echo pulse reaches the same size as the main pulse and therefore replaces it. This point labeled with a pink dot in Fig.~\ref{fig:DNS SNIPER}~(b) defines the pulse lifetime $T_{LT}$. Note that the observed instability resembles the satellite instability observed in the long cavity limit in the first principle delay algebraic models of vertical external-cavity surface-emitting semiconductor lasers  (VECSELs) and mode-locked integrated external-cavity surface-emitting lasers (MIXSELs)~\cite{SCM-PRL-19,SHJ-PRAp-20}. There, due to third-order dispersion from the Gires-Tournois-interferometer-like microcavity~\cite{GT-CRA-64}, pulses can have a series of decaying satellites on the leading edge. These satellites may become unstable, leading to a low-frequency modulation of the pulse envelope with a typical time scale of the order of hundreds of round trips. We note that in the short cavity case, a similar quasiperiodic scenario was found in \cite{Jaurigue2017,ludge2019semiconductor}. There,  a growth of the amplitude of the feedback-induced pulses with respect to the main pulse was observed as a function of the feedback strength $\eta$ and the cavity length $\tau_c$.
Note that the replacement process is not triggered by noise and it is inherent to the underlying deterministic system. As the result, the lifetime of the pulses is not arbitrary and one can see that the emerging solution is periodic with the fundamental period $T$ and the secondary period $T_{LT}$. Since the path-continuation of the torus orbits is not possible within DDE-BIFTOOL, extensive numerical simulations are conducted above the bifurcation point to track the solution in parameter space and to find the bifurcation point where the periodic orbit originates. These results are superposed with the corresponding TLS branch of the single pulse obtained from path-continuation and are presented in Fig.~\ref{fig:branch_sniper}~(a). Here, the blue line is obtained from the continuation while the red dots result from direct numerics and stand for the maximum intensity in each round-trip (cf. red circles in Fig.~\ref{fig:DNS SNIPER}). It can be seen that the periodic solution emerges from the fold point $F_s$ with a large period that depends on the distance to the bifurcation point $g_0-g_{0,cr}$; this indicates the global nature of this instability. To identify the type of the global bifurcation, the scaling of the oscillations period $T_{LT}$ close to the critical value $g_{0,cr}$ is analyzed. The resulting scaling is presented in Fig.~\ref{fig:branch_sniper}~(b),(c), where we show the evolution of $T_{LT}$ for two exemplary $(\alpha_g,\,\alpha_a)$ parameter sets and the characteristic scalings for saddle-node infinite-period (SNIPER) $(g_0-g_{0,cr})^{-1/2}$ (blue line) and homoclinic $-\ln(g_0-g_{0,cr})$ (green line) bifurcations as a function of the distance to the bifurcation point $g_{0,cr}$. The scaling is an inverted square root which indicates that the bifurcation of the SNIPER type. Note that since TLSs are periodic solutions, i.e., the observed bifurcation is a SNIPER bifurcation of \emph{periodic orbits}. Notice that for the short cavity case, the evolution of the period along the  quasiperiodic orbit induced by the feedback  was obtained as a function of the delay time (cf.  e.g., Fig. 9 (b) of  Ref. \cite{Jaurigue2017}). One can see, that the period diverges for one given point in the plane, spanned by the delay time and the delay strength. This may indicate the global nature of the bifurcation at this point, however, since no scaling analysis was shown, it is complicated to make any conclusion.
\section{Excitability} 
\begin{figure}
	\centering
	\includegraphics[width=1\columnwidth]{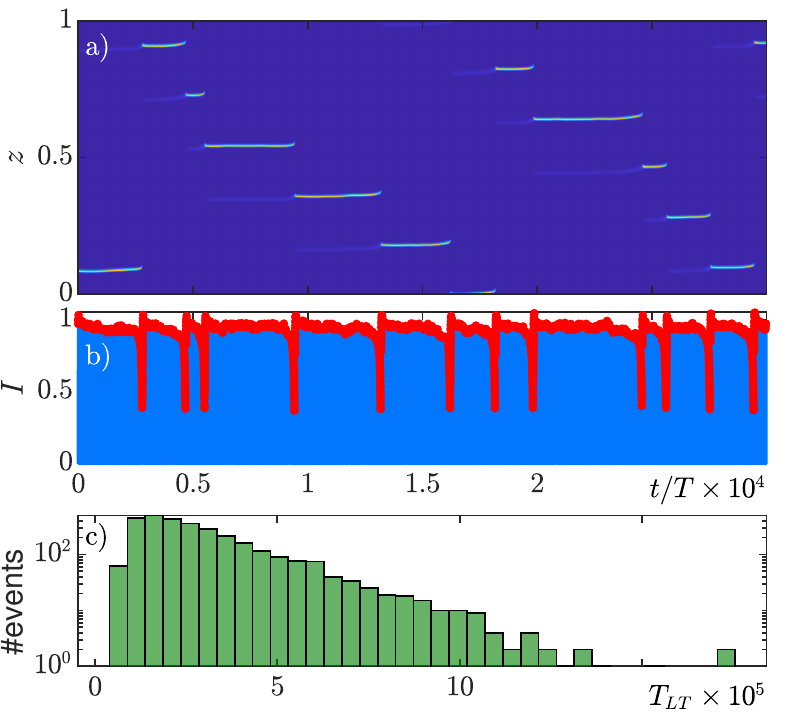}
	\caption{a) Pseudo-two-time representation and b) time trace of the single TLS for the parameter set  $(\alpha_g,\alpha_a,\eta,\tau_c,\tau_f,g)=(0.8,0.5,0.006,100,80,0.95)$ and noise of amplitude $\sigma=8\text{e}-3$ obtained from a direct numerical simulation of Eqs.~(\ref{eq:sys_eq_A})-(\ref{eq:sys_eq_Q}) using an Euler method. In b) the red circles denote the maximum in each round-trip. The distance between the minima of the red circles defines the lifetime $T_{LT}$ of each pulse. The statistical distribution of the different lifetimes is displayed in panel c) on a logarithmic scale for 3017 pulses.}
\label{fig:excitability_time_evol}
\end{figure}

Furthermore, the observed instability exhibits dynamical behavior characteristic of excitable systems. That is, it can be triggered by noise below the destabilizing bifurcation threshold. Indeed, close the bifurcation point, the influence of noise becomes important since it can help the weakly stable echo to explode. As a result, the period $T_{LT}$ can become highly irregular. Excitability plays  an important role in other laser systems considering the effects of TDF. Typically, they exhibit at their core mechanism an optical regeneration
process based upon excitability and can be  observed e.g., for  injected
lasers~\cite{GJT-NC-15,Munsberg2020},   excitable slow-fast Fitzhugh-Nagumo models~\cite{RAF-SR-16} or in the excitable laser with saturable absorber~\cite{SBB-PRL-14, Ruschel2020,terrien2021pulse} . There, an excitable system in combination with TDF gives rise to localized  states. In our case however, an excitable response is triggered by TDF in a system which already exhibits TLSs. We present in Fig.~\ref{fig:excitability_time_evol}~(a),(b) an exemplary pseudo-two-time diagram for a single pulse and the corresponding time trace in the excitable regime, respectively, under the influence of additive white Gaussian noise with a standard deviation of $\sigma$. The noise amplitude is chosen in a way to match typical fluctuations observed in experiments, in particular the works in \cite{CSV-OL-18,Bartolo2021}. Note that here an Euler method was used for the time integration as it more robust to noise compared to the Runge-Kutta method used for the other simulations. One can see the different lifetimes of the pulses as well as the fluctuations in the intensity as a result of the added noise. Moreover we found that in the presence of noise the lifetime is not determined by the intrinsic period but rather by an statistical distribution. In Fig. \ref{fig:excitability_time_evol}~(c), we show the statistical properties of the lifetimes of 3017 pulses under the influence of noise on a logarithmic scale. It can be seen that longer lifetimes are exponentially less likely to occur compared to shorter lifetimes. However, very short lifetimes do not occur as the replacement process takes a certain amount of time forming a lower border for $T_{LT}$. Note that the present lifetime distribution is similar to those found for the delay algebraic model of the MIXSEL~\cite{FG-PRE-20}. Finally, we note that above the bifurcation point, noise can also have the opposed effect to delay the next explosion which also results in fluctuation in the periodicity of the explosions.
\section{Two-parameter bifurcation analysis}

\begin{figure}
	\centering
	\includegraphics[width=1\columnwidth]{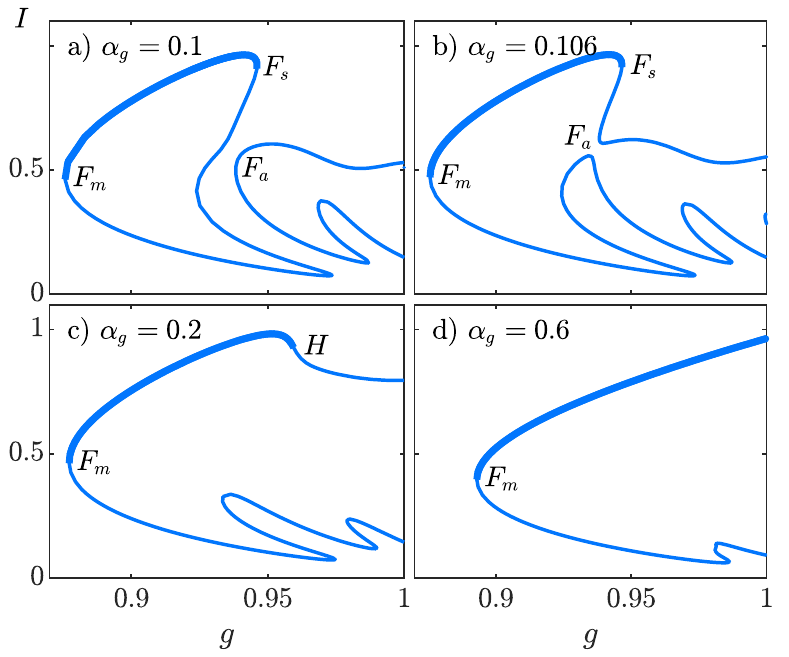}
	\caption{Branches of a single TLS of Eqs.~(\ref{eq:sys_eq_A})-(\ref{eq:sys_eq_Q}) for $\alpha_g=(0.1,\, 0.106,\, 0.2,\, 0.6)$ (a-d), respectively, showing the transition between the SNIPER bifurcation denoted by $F_s$ and a AH bifurcation marked by $H$. Thick and thin lines correspond to stable and unstable solutions, respectively. The value of $I=\max(|A|^2)$ is shown. $F_m$ and $F_a$ denote the folds of the main and the additional branches, respectively. Other parameters are $(\alpha_a,\eta,\tau_f)=(0,0.006,80)$.}
	\label{fig:bifdiag_alphaa_0_different_alphag}
\end{figure}

The results presented in Fig.~\ref{fig:branches_different_eta} indicate that not all TLS branches loose stability in a global SNIPER bifurcation but some do in a local AH bifurcation. This shows that the global instability only occurs for a certain parameter range. To clarify this issue we perform a two parameter bifurcation analysis. Here, we focus on the influence of the linewidth enhancement factors $\alpha_{a,g}$. For real devices, these values are nonzero and therefore we want to give an overview over a realistic parameter range. Figure~\ref{fig:bifdiag_alphaa_0_different_alphag} presents bifurcation diagrams of a FML solution in $(g, I)$ plane for different values of $\alpha_g$ while keeping $\alpha_a=0$. It can be seen that for $\alpha_g=0.1$ (panel (a)), another unstable branch with the main fold $F_a$ exists, which is not connected to the FML branch. If $\alpha_g$ is increased, this branch collides with the main TLS branch as showed in Fig.~\ref{fig:bifdiag_alphaa_0_different_alphag}~(b). After a further increase of $\alpha_g$, the SNIPER bifurcation point $F_s$ and the additional fold $F_a$ move towards each other and collide, see Fig.~\ref{fig:bifdiag_alphaa_0_different_alphag}~(c). Close to the collision, a AH bifurcation $H$ forms on the high-power part of the remaining branch. Finally, with growing $\alpha_g$ the Hopf bifurcation moves towards larger gain values (cf. Fig.~\ref{fig:bifdiag_alphaa_0_different_alphag}~(d)).

\begin{figure}
	\centering
	\includegraphics[width=1\columnwidth]{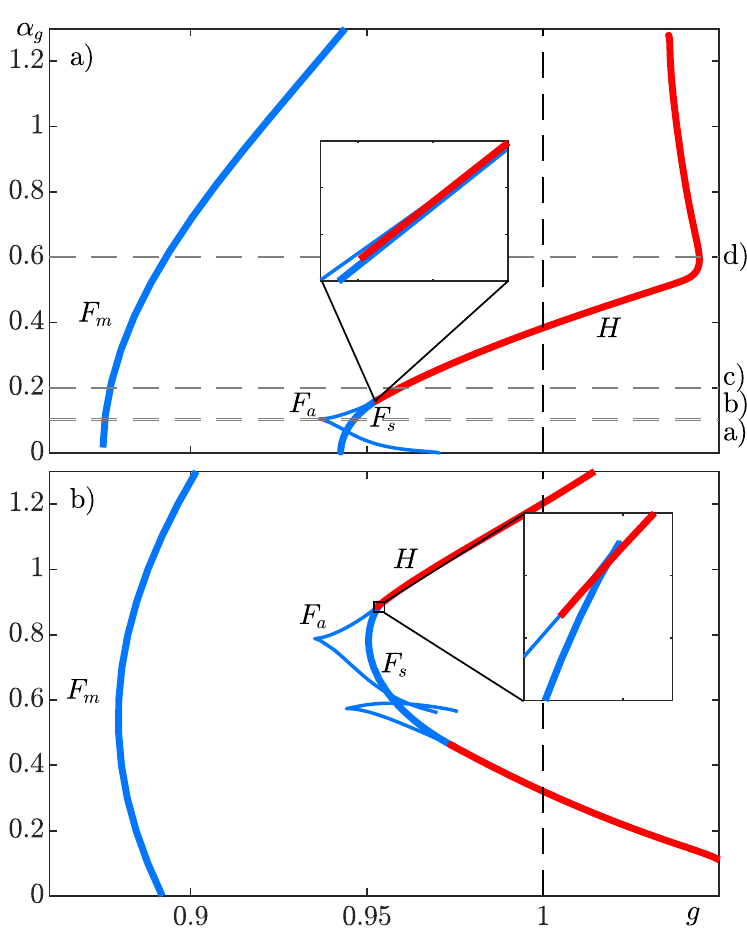}
	\caption{Two-parameter bifurcation diagrams in the ($g,\,\alpha_g$) plane of the single TLS of Eqs. (\ref{eq:sys_eq_A})-(\ref{eq:sys_eq_Q}) for a) $\alpha_a=0$ and b) $\alpha_a=0.5$, other parameters as in Fig.~\ref{fig:bifdiag_alphaa_0_different_alphag}. Blue and red lines stand for the thresholds of the saddle-node and AH bifurcations, respectively.  Thick and thin lines denote stable and unstable solutions, respectively. The horizontal dashed lines indicate the $\alpha_g$ values to obtain the branches in Fig.~\ref{fig:bifdiag_alphaa_0_different_alphag} (Note that a) and b) lie very close to each other). The insets give zoomed views on the cusp bifurcations region indicating that the AH bifurcation does not bifurcate from the tip of the cusp.}
	\label{fig:two_par_plane}
\end{figure}

In order to track this transition in details, a two parameter bifurcation analysis is performed. Figure \ref{fig:two_par_plane}~(a,b) shows the resulting bifurcation diagrams in the $(g,\,\alpha_g)$ plane for two different values of $\alpha_a$. Note that only the relevant bifurcations points $F_m, \,F_s,\, F_a$ as well as $H$ are tracked there because the dynamics becomes very complex especially on the unstable parts of the occurring  branches.

Figure~ \ref{fig:two_par_plane} ~(a) shows the bifurcation diagram for $\alpha_a=0$ that corresponds to the parameter set used for Fig.~\ref{fig:bifdiag_alphaa_0_different_alphag}. Thus, the dashed horizontal lines labeled with a)-d) correspond to the $\alpha_g$ values in Fig.~\ref{fig:bifdiag_alphaa_0_different_alphag}. One can see that for the small values of $\alpha_g$,  the fold $F_a$ of the unstable branch moves towards the  fold $F_s$ of the main branch and collides with it in a cusp.  The inset gives a zoomed view on the region of the cusp bifurcation and shows that the appearing periodic branch does not originate at the exact tip of the cusp but very close to it.  For increasing $\alpha_g$ the $H$ point moves towards the threshold. Hence, after a certain $\alpha_g$, the TLS is stable for the whole region between the main fold $F_m$ and $g_{0,th}$. Furthermore, one can observe that on the $F_a$ line there is another cusp bifurcation which lies exactly horizontal in the ($g,\,\alpha_g$)-plane and corresponds to the collision between the branches in Fig. \ref{fig:bifdiag_alphaa_0_different_alphag}~(a,b).
In Fig.~\ref{fig:two_par_plane}~(b), the bifurcation diagram in the ($g, \,\alpha_g$) plane is displayed for $\alpha_a=0.5$. The dynamics looks very similar to the case presented in Fig.~\ref{fig:two_par_plane}~(a) but the region of the SNIPER bifurcation $F_s$ is shifted towards larger values of $\alpha_g$ . Below and above this region the stability of the TLS is limited to the right by a AH bifurcation $H$ which originates in both cases close to the cusp of saddle-node bifurcations. This mechanism is completely analogue to the case $\alpha_a=0$. Therefore, one can conclude that varying $\alpha_a$ corresponds essentially to shifting the SNIPER bifurcation in the ($g,\,\alpha_g$) plane.
\section{Harmonic solutions and multistability}
We turn our attention to the influence of TDF on the harmonic solutions depicted in Fig.~\ref{fig:branch_harmonic}. These solutions consist of multiple equidistant pulses per round-trip. In the TLS regime, they coexist with the FML branch. Hence, the system is strongly multistable and the initial condition decides on which attractor the system settles.
\begin{figure}
	\centering
	\includegraphics[width=1\columnwidth]{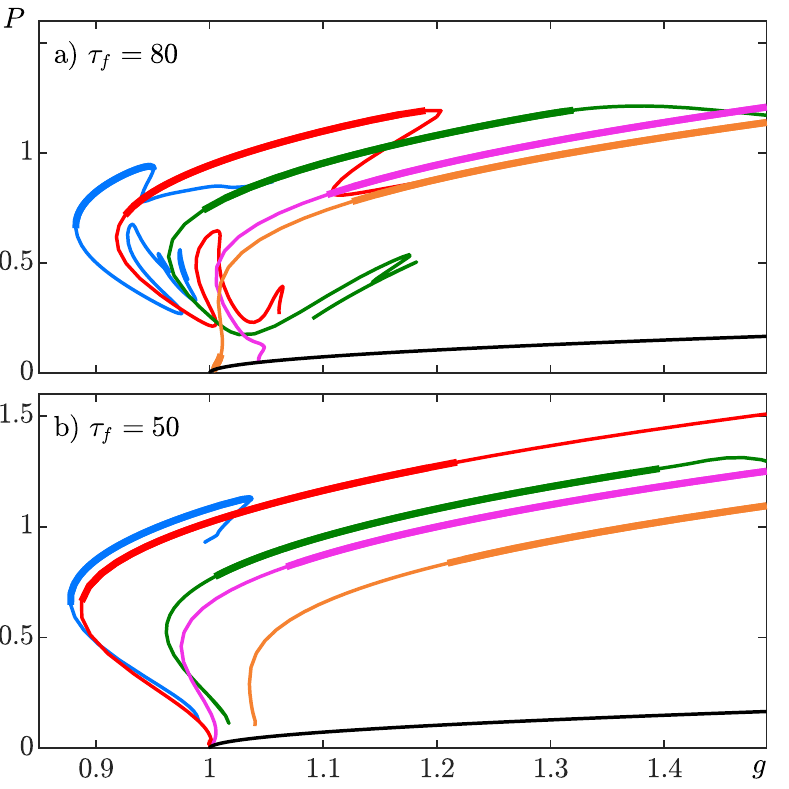}
	\caption{Bifurcation diagrams of Eqs.~ (\ref{eq:sys_eq_A})-(\ref{eq:sys_eq_Q}) in $g$ for the FML branch (blue) and harmonic HML$_{2-5}$ (red to yellow) solutions with two up to five pulses per round-trip for different values of $\tau_f$ which controls the location of the echo. The black line at the bottom corresponds to the CW branch. The parameters used are $(\alpha_g,\alpha_a,\eta)=(0.8,0.5,0.006)$. Resonant feedback increases the range of stability of the respective branches as seen in a) for the HML$_5$ branch, $\tau_f=80$ and in b) for the HML$_2$ branch, $\tau_f=50$. }
	\label{fig:branches_harmonic_different_tauf}
\end{figure}
In Fig.~\ref{fig:DNS} it was demonstrated that by choosing the feedback delay time $\tau_f$, the position of the echo $\Delta \tau$ can be directly controlled, cf. Ref.~\cite{Bartolo2021} and that its value
can either hinder or reinforce the stability of a given HML solution. To see the influence of the feedback on the stability of HML solutions, we compare  Fig.~\ref{fig:branches_harmonic_different_tauf}, where the branches of  the HML$_{2-5}$ solutions are displayed for two different feedback delay times $\tau_f$, with the results presented in  Fig.~\ref{fig:branch_harmonic} where the same parameters were used but feedback was turned off, i.e., $\eta=0$.
%In particular, Fig.~\ref{fig:branch_harmonic} shows that the absolute value of the electric field becomes smaller as more pulses are in the cavity. Next, the HML-branches gain stability in a AH bifurcation in contrast to the FML branch where the branch is stable after the fold.
Figure  \ref{fig:branches_harmonic_different_tauf}~(a) demonstrates that when introducing time delayed feedback with $\tau_f=80$ to the system,  the range of stability of the FML (blue) and HML$_{2,3,4}$ (red to magenta) is decreased, however, for the HML$_5$ branch (yellow) the range is increased. Indeed, for a cavity of length $\tau_c=100$ pulses are separated by $\tau_c/n=20$ for a HML$_5$ solution, which means that a feedback delay of $\tau_f=80$ creates a resonance between echo and pulse, i.e., the position of the echo $\Delta=\tau_c-\tau_f=20$ corresponds to the position of the previous pulse. The same mechanism can be observed for the HML$_2$ and HML$_4$ branches in Fig.~\ref{fig:branches_harmonic_different_tauf}~(b), where $\tau_f=50$. This demonstrates that feedback can have both a destabilizing and stabilizing effect depending on the pulse configuration in the cavity. This property can be exploited by using feedback as a solution selector as was recently shown experimentally in Ref.~\cite{Bartolo2021} .
\section{Pulse Dynamics around the Resonances}

\begin{figure*}
	\centering
	\includegraphics[width=1\textwidth]{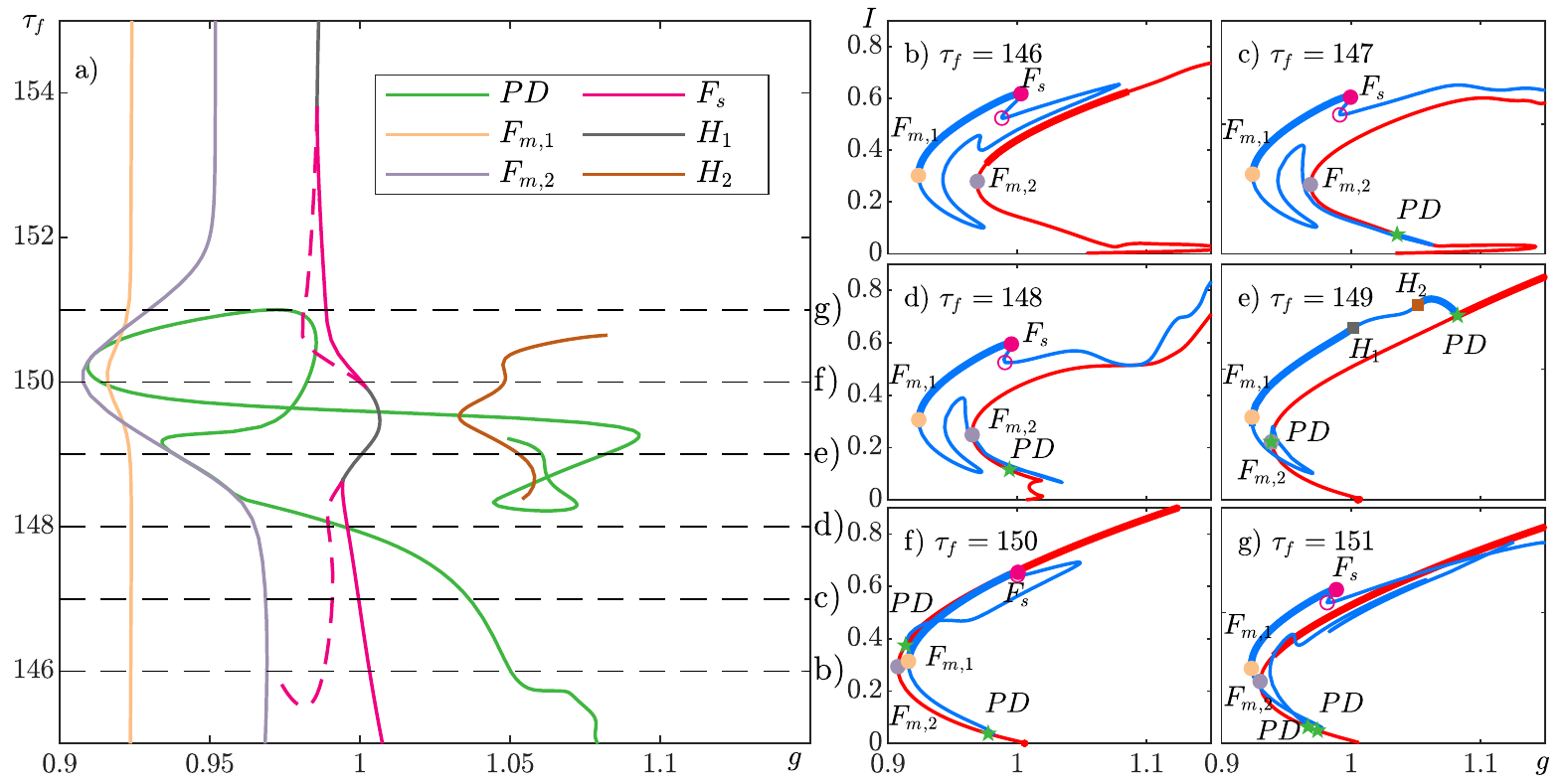}
	\caption{Bifurcation diagrams  of the FML and HML$_2$ solutions of Eqs.~(\ref{eq:sys_eq_A})-(\ref{eq:sys_eq_Q}). Panel a) shows the $(g,\, \tau_f)$ plane around the area of resonant feedback for $\tau_c=100$. The lines correspond to the threshold of a period doubling bifurcation ($PD$, green), the main folds of the FML (yellow) and HML$_2$ (light grey) branches ($F_{m,1},\,F_{m,2}$), respectively, the SNIPER bifurcation ($F_s$, magenta), and two AH bifurcations ($H_1,\,H_2$) in dark grey and brown, respectively. Note that some lines end abruptly due to numerical problems that occurred during the continuation. However, the parts not shown lay on unstable parts of the branches. The horizontal dashed lines indicate the cross section at which the bifurcation diagrams in panels b)-g) are plotted. In these panels the bifurcation points mentioned above are marked (circles for SN, squares for torus and stars for period-doubling bifurcations; colors according to panel a). The blue line stands for the FML branch, while the red line corresponds to the HML$_2$ branch. Thick and thin lines denote stable and unstable solutions, respectively. The  parameters  are $(\alpha_g,\alpha_a,\eta)=(1.5,0.5,0.009)$. The solution profiles along each branch in b)-g) can be found as an animation in the supplementary material .}
	\label{fig:two_par_run_tau2_flip}
\end{figure*}

The dynamics around the observed resonances is particularly rich as small changes in the echo's position decide between stabilizing or destabilizing the solution in question. If the echo appears just after the main pulse emission, the former is mostly harmless since the gain is already fully depleted by the main pulse, leading to the quasi-disappearance of the echo. On the contrary, if the echo occurs just before the main pulse emission, it is where the gain value is maximal, leading to a potential instability. We observe that this strong asymmetry is the result of the parity symmetry breaking induced by the carrier dynamics \cite{JCM-PRL-16}.
Further analysis revealed that the branches around the resonances are not independent but are connected via period doubling bifurcations. For a better understanding it is instructive to consider the $(g,\,\tau_f)$ plane in Fig.~\ref{fig:two_par_run_tau2_flip}~(a) where bifurcation thresholds of the FML and HML$_2$ branches are displayed for $\tau_c=100$. The corresponding bifurcation points for fixed values of $\tau_f$ are marked in Fig.~\ref{fig:two_par_run_tau2_flip}~(b)-(g). Note that some lines end abruptly due to numerical problems that occurred during the continuation. However, the parts not shown lay on unstable parts of the branches.
First, one can see the resonance appearing at $\tau_f=150$ as bumps in the fold lines $F_{m,1}$  of the FML (yellow) and $F_{m,2}$ of the HML$_2$ (light grey) branches in the $(g,\,\tau_f)$ plane. Here, the fold $F_{m,2}$ on the HML$_2$ branch is more affected because for the FML branch the resonance is only of the second order, i.e. the echo's echo. The corresponding bifurcation diagram in Fig.~\ref{fig:two_par_run_tau2_flip}~(f) shows a broad range of stability for both the FML (blue) and HML$_2$ (red) branches. Slightly below the resonance, at $\tau_f=149$ in Fig.~\ref{fig:two_par_run_tau2_flip}~(e), the HML$_2$ branch has a much smaller stability range while the FML branch remains less affected. However, its range of stability also becomes smaller since two AH bifurcations $H_1, H_2$ (dark grey, brown) arise on the branch.  For $\tau_f=148$ (cf. Fig.~\ref{fig:two_par_run_tau2_flip}~(d)) the HML$_2$ solution is completely unstable and only regains stability for $\tau_f=146$ when echo and pulse are separated sufficiently as shown in Fig.~\ref{fig:two_par_run_tau2_flip}~(b), while the FML stability is bounded between two fold points $F_{m,1}$ and the global bifurcation point $F_s$ (magenta).
Another important feature which emerges around the resonance is that the FML and HML$_2$ branches become connected by a period doubling (PD) bifurcation. In particular, first, as shown for for $\tau_f=146$ in Fig.~\ref{fig:two_par_run_tau2_flip}~(b), the FML and HML$_2$ branches are situated close to each other but are not connected.  However, at $\tau_f=147$, the FML branch touches the HML$_2$ branch on the low-power part and a PD bifurcation emerges, see Fig.~\ref{fig:two_par_run_tau2_flip}~(c). Here, the FML branch bifurcates from a PD point on the HML$_2$ branch just by doubling the period from $T\approx\tau_c/2$ to $T\approx\tau_c$. Thus, at this point, the solution profile consists of two pulses of same height. On the FML branch one of the pulses grows and the other becomes its echo while on the HML$_2$ branch both pulses remain the same height. For $\tau_f=149$, another  PD bifurcation also appears on the high power end of the branch, cf.~Fig.~\ref{fig:two_par_run_tau2_flip}~(e). In fact, it is the bifurcation in which the HML$_2$ branch gains stability. Finally, the PD bifurcation disappears for $\tau_f>151$. This can be seen in the  ($g,\,\tau_f$)-plane in Fig.~\ref{fig:two_par_run_tau2_flip}~(a). Interestingly, also a transition between the SNIPER and AH bifurcations can be seen around the resonance, where similar to Fig.~\ref{fig:two_par_plane} the AH bifurcation emerges close to a cusp point of saddle-node bifurcations (cf. solid and dashed magenta lines in Fig.~\ref{fig:two_par_run_tau2_flip}~(a) and magenta filled and open circles in the right panels (b)-(g)).
However, the brown and green lines do not cross as it is only in the chosen projection of the multidimensional parameter space they seem to do so.
We note that the observed dynamics around the resonances is similar to those observed for the short cavity~\cite{Jaurigue2017}.  Indeed, the effect of feedback
on periodic solutions is rather general as the ratio between $\tau_c$ and $\tau_f$ mostly defines the resulting dynamics. In particular, one can always consider the PML system as a nonlinear oscillator with natural period $\tau_{c}$. In this case, if an oscillator is subjected to delayed feedback, it make it so that one observes resonance tongues depending on the ratio between $\tau_{c}$ and $\tau_{f}$ (cf.~Ref.~\cite{Bartolo2021}, where this dependence was also studied experimentally).
\section{Conclusion}

In this paper the influence of the time-delayed optical feedback on the dynamics and stability of PML pulses in the long cavity regime was studied. This laser system is modeled by a system of delay differential equations, which includes delay terms associated with the laser cavity and the external feedback loop. We showed that TDF becomes visible in form of small echo pulses separated from each main pulse by $\Delta \tau=\tau_c-\tau_f$. The location of the echo can be controlled by the length of the cavity and its amplitude is determined by the feedback rate. It was demonstrated that when the echo is placed sufficiently close to the main pulse, it can destabilize the system when increasing the gain while increasing the gain value. 
There, the echos successively replace the main pulses after a certain pulse lifetime.

Analyzing the scaling behavior of the lifetime it is shown that the instability occurs in a SNIPER bifurcation of limit cycles. Note that a similar instability was described in Ref.~\cite{SHJ-PRAp-20} where the dynamics of passively mode-locked integrated external-cavity surface-emitting lasers were analyzed using a  dynamical model based upon delay algebraic equations. There, third order dispersion stemming from the lasing micro-cavity induces a train of decaying satellites which is similar to the effects of TDF discussed in this paper. Remarkably, the behavior in the two parameter bifurcation diagram (cf. Fig.~\ref{fig:two_par_plane}) is very similar in both systems which can be traced back to the similarity of the physical ingredients, a parasitic pulse rising on the leading edge of the main pulse. Furthermore, the influences of noise as well as the features of excitability were discussed.

Finally, the impact of the TDF on the harmonic solutions consisting of multiple pulses per round-trip was studied. When applying resonant feedback, i.e., the location of pulse coincide with a echo, the range of stability of this solution can be increased. It was also found, that around the resonance the pulses adapt their period, intensity and pulse width in order to include the echo in the pulse.
%JJ: I like that ! Why dont we put it !? SV: Sure , I think we forgot to coment out.
Furthermore, branches of different solutions are connected around the resonance by period doubling bifurcations. Note that this result match those obtained in Ref.~\cite{Jaurigue2017} for the harmonic solutions in the short cavity limit. However, we notice that in our system the value of the feedback is much smaller than in Ref.~\cite{Jaurigue2017}. This is due to the fact that, in the long cavity limit, the background is weakly stable, and particularly sensitive to perturbations. In general, the similarities and differences between the presented case of the long delay and the short cavity case, investigated in details in \cite{Jaurigue2017,ludge2019semiconductor} indicate that
it could be very instructive to study the transformation  of the complete bifurcation structure found in \cite{Jaurigue2017,ludge2019semiconductor} to the long cavity regime. However, it is out of the scope of this paper and will  be considered elsewhere.

More generally, the results presented here may serve as a guide for further analysis into the dynamics of pulsating systems with time-delayed feedback in the long cavity regime as we highlighted the susceptibility of this class of dynamical system to parasitic coherent re-injection.

\section*{Supplementary Material}
See supplementary material for animations showing the solution profiles along the various branches shown in Fig. \ref{fig:branches_different_eta} and Fig. \ref{fig:two_par_run_tau2_flip}.

\section*{Acknowledgments}
T.G.S. thanks the foundation “Studienstiftung des deutschen Volkes” for financial support, J.J. acknowledges the financial support of the MINECO Project MOVELIGHT (PGC2018-099637-B-100 AEI/FEDER UE), S.G. acknowledges the Universitat de les Illes Balears for funding a stay where part of this work was developed.

\section*{Data Availability}
The data that support the findings of this study are available from the corresponding author upon reasonable request.

\section*{References}
\bibliographystyle{apsrev4-2}
%\bibliographystyle{unsrt}
%\bibliography{full_new,extra}

%%%%%----------------- THE BIBLIOGRAPHY  ----------%%%%%

%apsrev4-2.bst 2019-01-14 (MD) hand-edited version of apsrev4-1.bst
%Control: key (0)
%Control: author (72) initials jnrlst
%Control: editor formatted (1) identically to author
%Control: production of article title (-1) disabled
%Control: page (0) single
%Control: year (1) truncated
%Control: production of eprint (0) enabled
%

\end{document}